\documentclass[a4]{PoS}

\usepackage{bm}
\usepackage{amsmath}

\newcommand{\mrm}{\mathrm}
\newcommand{\mc}{\mathcal}
\newcommand{\rk}{R_{K^{(*)}}}
\newcommand{\rd}{R_{D^{(*)}}}

\newcommand {\e}[1]{\mathrm{~#1}}       
\newcommand {\re}[0]{\mathrm{Re}}

\title{Beyond the Standard Model, guided by Lepton Universality}

\ShortTitle{BSM from LFU}

\author{\speaker{Nejc Ko\v snik}\\
        Jo\v zef Stefan Institute and Faculty of Mathematics and Physics, University of Ljubljana \\
        E-mail: \email{nejc.kosnik@ijs.si}}

\abstract{We discuss the current status of lepton flavour universality observables in semileptonic charged and neutral current $B$ meson decays. We recapitulate the current experimental situation. Assuming the current tensions with the SM are of beyond the Standard Model origin we present their implications for the effective theory couplings. Then we discuss general leptoquark models before finally focussing on the model with two scalar leptoquarks ($R_2$ and $S_3$) which have couplings connected by the underlying Grand Unified Theory framework. We present the viability of the model with regard to lepton flavour universality and present opportunities to search for this model in future high-intensity and high-energy experiments.}

\FullConference{18th International Conference on B-Physics at Frontier Machines - Beauty2019 -\\
		29 September / 4 October, 2019\\
		Ljubljana, Slovenia}

\begin{document}

\section{Introduction}
The Standard Model, our best theory describing the basic constituents and their interactions, has passed so far all the challenges posed by the numerous tests performed at high-energy and high-intensity experiments. The flavour sector of the Standard Model~(SM) is the richest in number of parameters describing fermion masses and mixing parameters among them. The flavour and CP-violating effects have been confirmed by the $B$-factories to conform to the Cabibbo-Kobayashi-Maskawa~(CKM) framework. We do know SM is superseeded by some BSM model due to several shortcomings: unexplained neutrino masses, breakdown of the theory at the Planck scale, and possibly the lack of dark matter particle. There are also conceptual issues such as the electroweak hierarchy problem and puzzling hierarchies in masses and mixings. The measurements at future $B$-meson experiments can further challenge the flavour sector by making careful tests of the SM predictions.

The observables that test lepton flavour universality~(LFU) have been essential in SM validation. Violation of a large flavour symmetry of the SM is confined to the Yukawa sector. According to the SM pattern of Yukawas lepton flavours are conserved and are treated universally by gauge bosons. Only Higgs Yukawas and masses separate charged lepton flavours.
In LFU ratios we compare decay widths or cross sections which differ only in charged lepton flavour. The gauge and CKM factors cancel out in the ratio, whereas hadronic physics parameters entering processes, such as decay constants and form factors, cancel in processes with two body final states and the LFU ratio can be expressed as a function of particle masses. In processes with $(\geq 3)$-body kinematics the dependence on form factors is partially cancelled and vanishes in the limit where lepton masses can be neglected.

The persistent hints of violation of the LFU predictions in the SM have been reported in semileptonic $B$ meson decays. In charged-current decay $B \to D^{(*)} l \nu$ the two ratios $\rd = \mc{B}(B \to D^{(*)}) \tau \nu)/\mc{B}(B \to D^{(*)}) \ell \nu)$, with $\ell = e,\mu$ have been measured by BaBar, Belle, while LHCb experiment measured $R_{D^*}$. The current world average~(WA) of $\rd$ measurements by HFLAV~\cite{Amhis:2019ckw} is based on experimental results~\cite{1205.5442,1303.0571,1507.03233,1506.08614,1612.00529,1709.00129,1708.08856,1709.02505,1904.08794} and exhibits more than $3\sigma$ tension of SM with the data:
\begin{equation}
  \label{eq:RDexp}
\left.  \begin{array}{rcl}
  R_D^\mrm{WA} & = & 0.340(27)(13) \\
    R_{D^*}^\mrm{WA} &= & 0.295(11)(8)
        \end{array}
      \right\} \,\,
      \mrm{Corr}(R_D^\mrm{WA}, R_{D^*}^\mrm{WA}) = -0.38,\qquad
      \begin{array}{rcl}
        R_D^\mrm{SM} & = & 0.299(3), \\
        R_{D^*}^\mrm{SM} &= & 0.258(5).
        \end{array}
      \end{equation}
      The above SM predictions refer to HFLAV average of results in~\cite{1606.08030,1703.05330,1707.09509,1707.09977}. Here the
      large mass of the $\tau$ implies large LFU violation already in
      the SM. In rare semileptonic decays $B \to K \ell \ell$ where
      $\ell = e,\mu$ the SM contribution should respect LFU at
      $q^2 \gtrsim m_\mu^2$ and the ratios
      $\rk = \mc{B}(B \to K^{(*)} \mu^+ \mu^-)/ \mc{B}(B \to
      K^{(*)} e^+ e^-)$ are expected to be close to $1$. The LHCb experiment measured $\rk$ at low invariant mass bin $[1.1, 6.0]\e{GeV}^2$ of $q^2$,  $R_{K^*}$ was measured also in the ultra low bin $[0.045,1.1]\e{GeV^2}$:
        \begin{align}          
        \phantom{aaa}R_K^\mrm{LHCb}\big|_{[1.1,6]\e{GeV}^2} &= 0.846 ^{+0.060+0.016}_{-0.054-0.014}~\cite{Aaij:2019wad}, \qquad  R_K^\mrm{th}\big|_{[1.1,6]\e{GeV}^2} = 1.00\pm 0.01~\cite{1605.07633},\\
        R_{K^*}^\mrm{LHCb}\big|_{[1.1,6]\e{GeV}^2} &= 0.69^{+0.11}_{-0.07}\pm 0.03~\cite{Aaij:2017vbb}, \qquad R_{K^*}^\mrm{th}\big|_{[1.1,6]\e{GeV}^2} = 1.00\pm 0.01 ~\cite{1605.07633},\\
        R_{K^*}^\mrm{LHCb}\big|_{<1.1\e{GeV}^2} &= 0.66^{+0.11}_{-0.07}\pm 0.03~\cite{Aaij:2017vbb}, \qquad  R_{K^*}^\mrm{th}\big|_{<1.1\e{GeV}^2} = 0.983 \pm 0.014 ~\cite{1605.07633}.
      \end{align}
      The two experimental errors are split into statistical and
      systematic one, and are also given in that order. All three
      results have systematic errors well under control and are
      consistently below the SM predictions.  The combined
      significance of about $4\sigma$.

      The two LFU ratios are currently our only potential hints of
      flavour effects beyond the SM. In this talk we explore what are
      the implications of lepton flavour universality hints for BSM
      scenarios.

      \section{Effective theory view}
      The weak effective Lagrangian for semileptonic decays is
      an appropriate framework to parameterize general BSM effects in
      $\rd$. At renormalization scale $\mu = m_b$ the relevant interactions for $b \to c \tau \bar \nu_\tau$ transitions are:
      \begin{equation}        \label{eq:1}
        \begin{split}          
       \mc{L}_\mrm{SL} = \frac{4G_F V_{cb}}{\sqrt{2}} \Big[(1&+g_{V_L}) (\bar c_L \gamma^\mu b_L) (\bar\tau_L \gamma_\mu \nu_{\tau L}) + g_{V_R} (\bar c_R \gamma^\mu b_R) (\bar\tau_L \gamma_\mu \nu_{\tau L}) \\
        &+g_{S_L} (\bar c_R b_L) (\bar\tau_R \nu_{\tau L}) + g_{S_R} (\bar c_L b_R) (\bar\tau_R \nu_{\tau L}) +g_{T} (\bar c_R \sigma_{\mu\nu} b_L) (\bar\tau_R
        \sigma^{\mu\nu} \nu_{\tau L})\Big].
              \end{split}
            \end{equation}
            Here we will assume that BSM in $\rd$ does not contribute
            to $b\to c \ell \bar\nu_\ell$.  Using the form factors
            from the lattice~\cite{1505.03925,1503.07237,1311.5071},
            and using further theoretical~\cite{1703.05330} and
            experimental results~\cite{1612.07233} the authors of
            Ref.~\cite{1806.10155} presented compact numerical
            expressions for $\rd$:
            \begin{equation}
              \begin{split}               
                \frac{\rd}{R_{D^{(*)}}^\mrm{SM}} = &1 +  a_S^{D^{(*)}} |g_{S_R} + g_{S_L}|^2 + a_P^{D^{(*)}} |g_{S_R} - g_{S_L}|^2 + a_T^{D^{(*)}} |g_T|^2\\
                                &+ a_{S V_L}^{D^{(*)}} \re[g_{S_R} + g_{S_L}] +  a_{P V_L}^{D^{(*)}} \re[g_{S_R} - g_{S_L}] + a_{T V_L}^{D^{(*)}} \re[g_T].
              \end{split}
            \end{equation}
            Values of coefficients $a_i^{D^{(*)}}$ can be found in~\cite{1806.10155} or~\cite{1811.09603}. Using these coefficents it was found in Ref.~\cite{1808.08179} that the tensor (real $g_T$) scenario fits the data best, followed by the left-handed scenario (real $g_{V_L}$). The same authors also studied various leptoquark~(LQ) scenarios with a combination of scalar and tensor coupling and found two LQ scenarios that accommodate the data: real $g_{S_L} = -4 g_T$, or imaginary  $g_{S_L} = 4 g_T$, where both relations hold at the LQ scale and are modified by QCD renormalization at the scale $m_b$. Additional information is available in the $q^2$ shapes of decay spectra as shown by Ref.~\cite{1506.08896}. Belle experiment also measured in $B \to D^* \tau \nu$ the polarization of $\tau$ and longitudinal polarization~\cite{1903.03102} of $D^*$ and the latter observable $F_L^{D^*,\mrm{Belle}} = 0.60 \pm 0.08(\mrm{stat}) \pm 0.04(\mrm{sys})$ is somewhat below the SM value $0.46\pm 0.04$~\cite{1606.03164} as well as below most of BSM models that fit $\rd$ well~\cite{1606.03164}. Furthermore, it was shown that various LQ models could be separated one from another by more preciese measurement of the $\tau$ polarization~\cite{1811.08899}. Scalar interactions $g_{S_L}$, $g_{S_R}$ could also severely disturb the branching fraction $B_c \to \tau \nu$. The latter can be constrained from the LEP data~\cite{1611.06676, 1708.04072} and together with the known $B_c$ lifetime provide relevant constraint on $g_{S_R} - g_{S_L}$~\cite{1605.09308,1811.09603}. Naive scale of NP from $\rd$ can be inferred to be few TeV if the effective interactions (e.g. $g_{V_L}$) are assumed to be of order 1. Such large BSM effects might also generate dangerous effects via radiative corrections in precisely measured lepton decays, see e.g.~\cite{1606.00524,1705.00929}.

            The $\rk$ anomalies and related measurements in $b \to s \mu^+ \mu^-$ transitions are described at low energies with
            \begin{equation}
              \label{eq:Leff-RK}
              \mc{L}_{b \to s \ell\ell} = \frac{4G_F}{\sqrt{2}}V_{tb} V_{ts}^* \sum_{\substack{i=7,9,10,S,P\\\phantom{i=}9',10',S',P'}} C_i 
\mathcal{O}_i,
            \end{equation}
            where the most relevant operators for the anomalies are
            \begin{equation}
              \label{eq:4}
              \begin{split}                
              \mathcal{O}_9 &= \frac{e^2}{16\pi^2} (\bar s_L \gamma^\mu b_L) (\bar \mu \gamma_\mu \mu),\\
\mathcal{O}_{10} &= \frac{e^2}{16\pi^2} (\bar s_L \gamma^\mu b_L) (\bar \mu \gamma_\mu \gamma^5 \mu),\\
\mathcal{O}_{S} &= \frac{e^2}{16\pi^2} (\bar s_L b_R) (\bar \mu \mu),\\
\mathcal{O}_{P} &= \frac{e^2}{16\pi^2} (\bar s_L b_R) (\bar \mu \gamma^5 \mu).
\end{split}
\end{equation}
The fits of $\rk$ indicate that left-handed scenario, $C_9 = -C_{10}$, gives a very good description of the data, where we have assumed SM-like couplings to the electrons. Furthermore, if we include many available observables of the $b\to s\ell \ell$ transitions from Belle, BaBar, and LHCb in the global fit we find that NP couplings to electrons are indeed not necessary, whereas there is strong indication for non-zero $C_9$~\cite{1903.09578}. The scale of generic NP model with order 1 flavour changing neutral current couplings to explain the $\rk$ is few $10$'s of TeV. The requirement of perturbative unitarity in $qq \to \ell \ell$ scattering give the upper bound on the scale of $\rk$ to be $\lesssim 10\e{TeV}$ and $\lesssim 100\e{TeV}$ for BSM explaining $\rd$~\cite{1706.01868}.

\section{Leptoquark models}
Here we focus on the LQ mediators with suitable properties for $\rd$ and/or $\rk$. 
The phenomenological advantage of LQs is that their natural tree-level contribution are semi-leptonic processes whereas their SM charges do not allow for tree-level contributions to neutral meson mixing amplitudes. The latter are major obstacle to $Z'$ models that address $\rk$. The recent phenomenological evaluation of LQs with respect to their role in LFU observables, flavour constraints, $Z$-pole observables, high-$p_T$ constraints was undertaken in~\cite{1808.08179} and confirmed that vector leptoquark $U_1$ in the representation $(3,1,2/3)$ is the only LQ that explains all observed LFU. The $U_1$ has been proposed before~\cite{1706.07808} and was later embedded in a necessary UV completion, based on Pati-Salam unified groups for each generation~\cite{1712.01368,1805.09328,1903.11517} or in the context of 4321 model~\cite{1802.04274}, among others. Ref.~\cite{1808.08179} found no appropriate scalar LQ for both LFU anomalies. $S_3$ with purely left-handed couplings can accommodate $\rk$ while $R_2$ and $S_1$ are suitable for $\rd$. At loop level $R_2$ has the desired Lorentz structure of couplings to explain $\rk$ but the needed couplings are in tension with LEP and LHC constraints~\cite{1704.05835,1805.04917}. Here we focus on scenarios with two scalar LQs. One possibility is to take $S_1$ and $S_3$ which is the route described in ~\cite{1703.09226} while here we will entertain the possibility to employ pair of $R_2$ and $S_3$ LQs.
    
\section{$R_2$ and $S_3$ from Grand Unified Theory}
The Yukawa couplings of $R_2$ and $S_3$ with the quarks and leptons in the mass basis can be written as~\cite{Becirevic:2018afm}
\begin{align}
\label{eq:two}
\begin{split}
\mathcal{L}_\mrm{Yuk} = 
&+(V Y_R E_R^\dagger)^{ij} \bar{u}_{Li}\ell_{Rj}R_2^{\frac{5}{3}} + (Y_R E_R^\dagger)^{ij} \bar{d}_{Li}\ell_{Rj} R_2^{\frac{2}{3}}
+(U_R Y_L U)^{ij} \bar{u}_{Ri} \nu_{Lj} R_2^{\frac{2}{3}}\\
&- (U_R Y_L)^{ij} \bar{u}_{Ri}\ell_{Lj} R_2^{\frac{5}{3}}+(Y_L U)^{ij} \bar{d}^C_{Li} \nu_{Lj} S_3^{\frac{1}{3}} - \sqrt{2}(V^* Y_L U)^{ij} \bar{u}^C_{Li} \nu_{Lj} S_3^{-\frac{2}{3}}\\
&+\sqrt{2} Y_L^{ij} \bar{d}^C_{Li} \ell_{Lj} S_3^{\frac{4}{3}} +(V^* Y_L)^{ij} \bar{u}^C_{Li} \ell_{Lj} S_3^{\frac{1}{3}}.
\end{split} 
\end{align}
Here $Y_L$, $Y_R$ are the arbitrary LQ Yukawa matrices, $R_2^{(Q)}$
and $S_3^{(Q)}$ are LQ charge eigenstates of LQs. The unitary matrices
$U_{L,R}$, $D_{L,R}$, $E_{L,R}$, and $N_L$ rotate between mass and
gauge basis of up-type quarks, down-type quarks, charged leptons, and
neutrinos, respectively. $V \equiv U_L D_L^\dagger = U_L$ is the CKM
matrix, $U\equiv E_L N_L^\dagger = N_L^\dagger$ is the PMNS matrix.
The following numerical pattern is assumed for the Yukawa matrices:
\begin{equation}
\label{eq:yL-yR}
Y_R E_R^\dagger = \begin{pmatrix}
0 & 0 & 0\\ 
0 & 0 & 0\\ 
0 & 0 & y_R^{b\tau}
\end{pmatrix},\qquad
U_R Y_L = \begin{pmatrix}
0 & 0 & 0\\ 
0 & y_L^{c\mu} & y_L^{c\tau}\\ 
0 & 0 & 0
\end{pmatrix},
\end{equation}
where $U_R$ is a rotation by angle $\theta$ between second and third
generation. The parameters of the model are thus $m_{R_2}$, $m_{S_3}$,
$y_R^{b\tau}$, $y_L^{c\mu}$, $y_L^{c\tau}$, and $\theta$. In a low
energy LQ setting there is no reason for the Yukawa couplings of $S_3$
to be related to the ones of $R_2$. In our case we consider two
leptoquarks within the $SU(5)$ based unification model where the
scalar sector contains representations of dimension $\bm{45}$ and
$\bm{50}$. The SM fermions reside in $\overline{\bm{5}}_{i}$ and
$\bm{10}_i$, with $i(=1,2,3)$ counting generations. All the low-energy
operators of Eq.~\eqref{eq:two} can be generated with $SU(5)$
contractions
$a^{ij} \bm{10}_i \overline{\bm{5}}_j \overline{\bm{45}}$, and
$b^{ij} \bm{10}_i \bm{10}_j \bm{50}$, where $a$ and $b(=b^T)$ are
matrices in the flavour space. The former contraction couples
$R_2 \in \bm{45}$ ($S_3 \in \bm{45}$) with the right-handed up-type
quarks (quark doublets) and leptonic doublets, while the latter
generates couplings of $R_2 \in \bm{50}$ with the quark doublets and
right-handed charged leptons. To break $SU(5)$ down to the SM we use
scalar representation $\bm{24}$ and write a term in the scalar
potential $m\, \bm{45} \, \overline{\bm{50}} \, \bm{24}$. The two
$R_2$ leptoquarks that reside in $\bm{45}$ and $\bm{50}$ then mix and
allow us to have one light $R_2$ and one heavy $R_2$ in the spectrum,
where the latter state completely decouples from the low-energy
spectrum for large values of $m$~\cite{Becirevic:2018afm}.  LQs can be
dangerous for proton decay if they couple to diquarks. The $S_3$
leptoquark would not couple to the diquark if $SU(5)$ contraction
$c^{ij} \bm{10}_i \bm{10}_j \bm{45}$ was forbidden or
suppressed. Furthermore, $S_3$ must not mix with any other LQ with
diquark couplings. Both conditions can be met in a generic $SU(5)$
framework~\cite{Dorsner:2017wwn}.

At high scale $\Lambda = m_{R_2}$ the Wilson coefficients for the charged current processes are:
\begin{equation}
\label{eq:semilep-WC}
\begin{split}
g_{S_L}(\Lambda) = 4 \, g_T(\Lambda) &= \frac{y_L^{c\tau}\, {y_R^{b\tau}}^{\ast}}{4 \sqrt{2} \, m_{R_2}^2 \, G_F V_{cb}}.
\end{split}  
\end{equation}
This is the only flavour of charged current semileptonic process affected by $R_2$.
The $\rk$ anomaly is accounted for through left-handed tree-level contributions
of $S_3$ to the vector and axial-vector Wilson coefficients~\cite{Dorsner:2017ufx}
%
\begin{equation}
\begin{split}
  \delta C_9 = - \delta C_{10}  &=\dfrac{\pi v^2}{ \lambda_t \alpha_{\mathrm{em}}} \frac{\sin 2\theta\, (y_L^{c\mu})^2}{2 m_{S_3}^2}.
\end{split}
\end{equation}
Here the mixing angle enters as $\sin 2\theta$, originating from the matrix $U_R$, and plays an important role in suppressing effect in $R_{K^{(*)}}$ relative to the one in $R_{D^{(*)}}$.
The $1\,\sigma$ interval  $C_9=-C_{10} \in (-0.85,-0.50)$ has been obtained by performing a fit to $R_K$, $R_{K^{\ast}}$, and $\mathcal{B}(B_s\to \mu\mu)$.
The left-handed (weak triplet) nature of the $S_3$ LQ imply contributions to both neutral and charged current semileptonic processes. Among the charged current observables the LFU ratios $R_{D^{(\ast)}}^{\mu/e} = \mathcal{B}(B\to D^{(\ast)}\mu \bar{\nu})/\mathcal{B}(B\to D^{(\ast)} e \bar{\nu})$ impose severe constraints on $S_3$ couplings. We have also considered $\mathcal{B}(B\to \tau \bar{\nu})$ and the kaon LFU ratio $R^K_{e/\mu}= \Gamma(K^-\to e^- \bar{\nu})/\Gamma(K^-\to \mu^- \bar{\nu})$. 
The $b\to s \nu\bar \nu$ constraints are not taken as inputs as constraints. Instead we predict $R_{\nu\nu}^{(\ast)}$ and compare it to experimental bounds, $R_{\nu\nu}<3.9$ and $R_{\nu\nu}^{\ast}<2.7$~\cite{Grygier:2017tzo}. The loop-induced neutral-current constraints affect both LQ's couplings. 
We have taken into account the $B_s-\bar{B}_s$ mixing frequency, which is modified by the $S_3$ box-diagram, proportional to 
$\sin^2 2\theta \left[(y_L^{c\mu})^2 + (y_L^{c\tau})^2\right]^2/m_{S_3}^2$,
and the upper limit on lepton flavour violating $\tau$ decays $\mathcal{B}(\tau\to\mu\phi)$, $\mathcal{B}(\tau\to\mu\gamma)$. The $Z$-boson couplings to leptons measured at
LEP~\cite{ALEPH:2005ab} are also modified at loop level by both LQs.
\begin{figure}[!hb]
  \centering
  \includegraphics[scale=0.5]{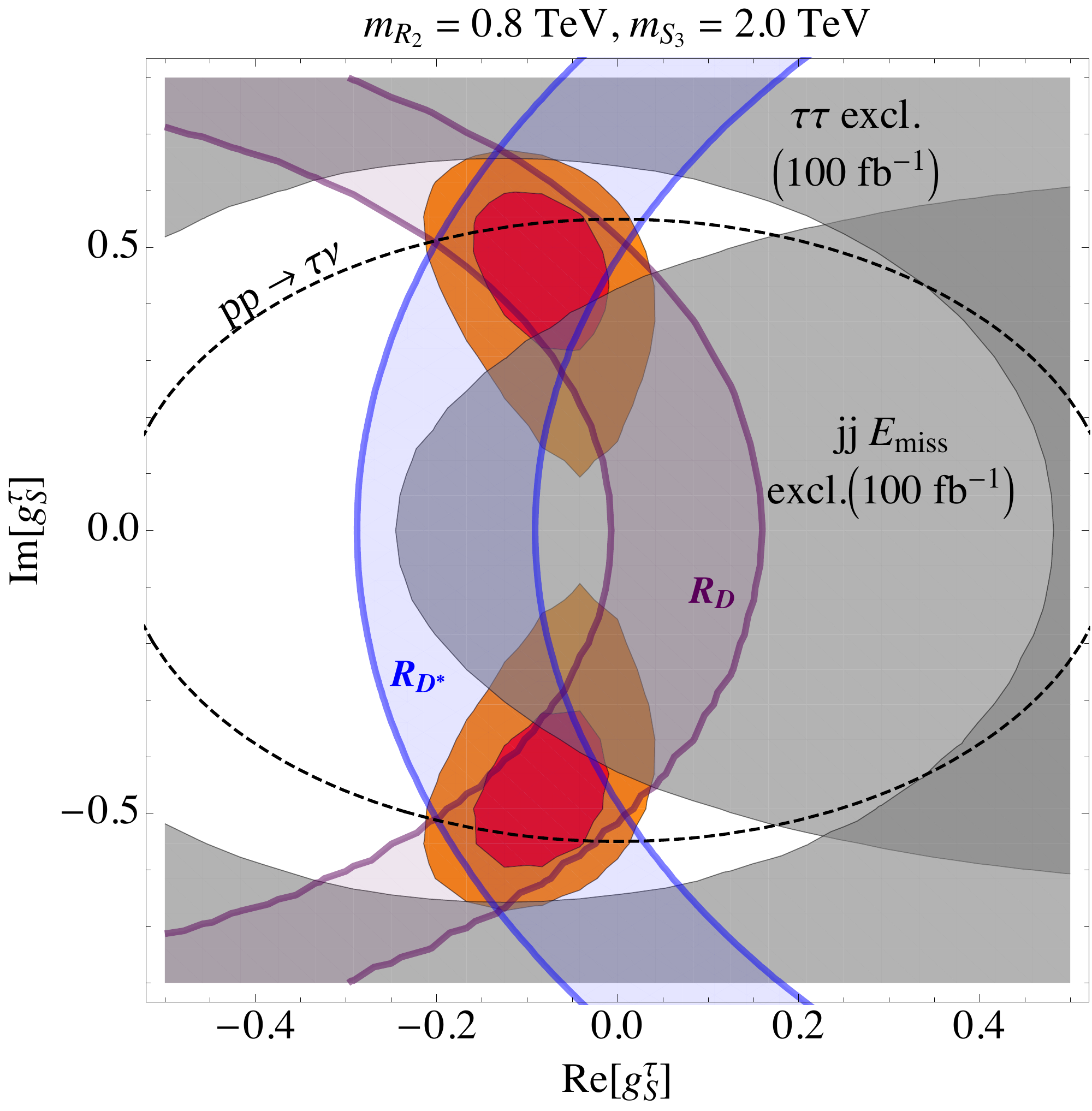}
  \caption{Results of the flavour fit in the $g_{S_L}$ plane, scalar coupling entering  the transition $b\to c\tau \bar{\nu}_\tau$. The $1\,\sigma (2\,\sigma)$ fitted regions are rendered in red (orange). Separate constraints from $R_D$ and $R_{D^\ast}$ to $2\,\sigma$ accuracy are shown by the blue and purple regions, respectively. The LHC exclusions are depicted by the gray regions. Dashed circle denotes the $pp \to \tau \nu$ constraint.}
    \label{fig:gSplotRDst}
\end{figure}

Taking into account the aforementioned flavour observables we have performed a fit for parameters $y_R^{b\tau}$, $y_L^{c\mu}$, $y_L^{c\tau}$ and $\theta$, while fixing the masses to $m_{R_2} = 0.8\e{TeV}$ and $m_{S_3} = 2\e{TeV}$. 
The opposite sign of interference terms in $R_D$ and $R_{D^*}$ require complex Wilson coefficient $g_{S_L}$ (Fig.~\ref{fig:gSplotRDst}), where we have put the complex phase in $y^R_{b\tau}$. The SM is excluded at $3.6\,\sigma$, the best fit point provides a good agreement with $R_{D^{(\ast)}}$ and $R_{K^{(\ast)}}$. Note that the required large imaginary part in $y_R^{b\tau}$ could be in principle tested in the future experiments measuring neutron EDM~\cite{Dekens:2018bci}. The best fit point is consistent with the LHC constraints~\cite{Becirevic:2018afm} superimposed in gray on the same plot. The $pp \to \tau \nu$ constraint in the effective theory approximation excludes the region outside dashed circle~\cite{1811.07920} in Fig.~\ref{fig:gSplotRDst}.

\begin{figure}[!htbp]
  \centering\includegraphics[scale=.7]{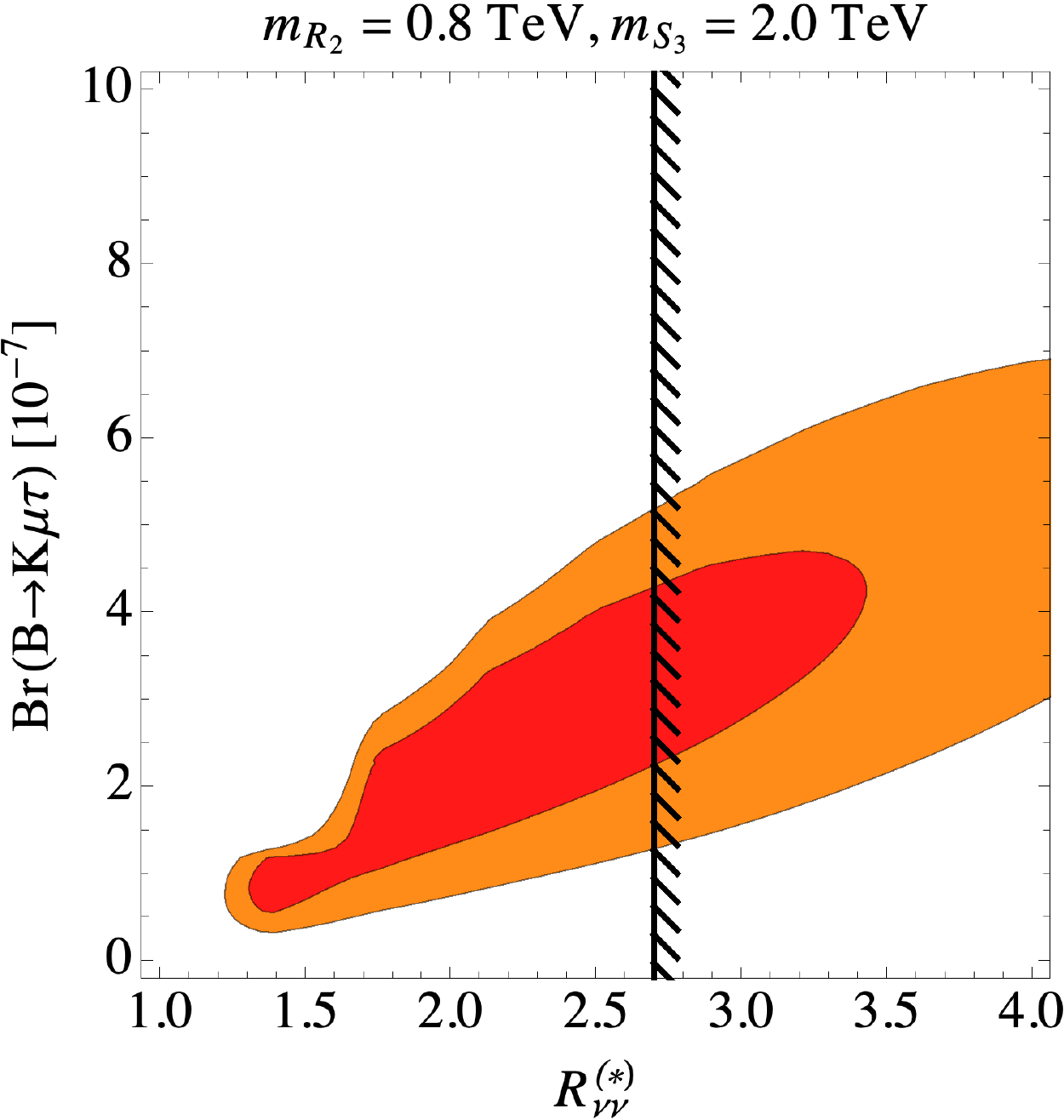}
  \caption{Predicted $\mathcal{B}(B\to K \mu \tau)$ is plotted against predicted $R_{\nu\nu}=\mathcal{B}(B\to K^{(\ast)} \nu \bar{\nu})/\mathcal{B}(B\to K^{(\ast)} \nu \bar{\nu})^{\mathrm{SM}}$ for the $1\,\sigma$ (red) and $2\,\sigma$ (orange) regions of Fig.~\ref{fig:gSplotRDst}. The black line denotes the current experimental limit, $R_{\nu\nu}^{\ast}<2.7$~\cite{Grygier:2017tzo}.}
  \label{fig:prediction}
\end{figure}

\noindent The consistency of model with low energy data requires that $\mathcal{B}(B\to K\mu\tau)$ is bounded and 
at $1\,\sigma$ we obtain $1.1 \times 10^{-7}\lesssim  \mathcal{B}( B\to K \mu^\pm \tau^\mp) \lesssim 6.5\times 10^{-7}$, whereas related decay modes are predicted to be $\mathcal{B}(B\to K^\ast \mu\tau)\approx 1.9\times \mathcal{B}(B\to K \mu \tau)$ and $\mathcal{B}(B_s\to \mu\tau)\approx  0.9 \times \mathcal{B}(B\to K \mu \tau)$. Another important prediction is a $\gtrsim 50\%$ enhancement of $\mathcal{B}(B\to K ^{(\ast)} \nu \nu)$, which will be further tested at Belle 2. Remarkably, these two observables are highly correlated~(Fig.~\ref{fig:prediction}). Furthermore, we derive a lower bound on $\mathcal{B}(\tau \to \mu \gamma)$, just below the current experimental limit,  $\mathcal{B}(\tau \to \mu \gamma) \gtrsim 1.5\times 10^{-8}$.

Finally, our description of $R_{D^{(\ast)}}$ anomaly requires relatively light LQ states, not far from the TeV scale, and these LQs are necessarily accessible at the LHC, as we discuss in Ref.~\cite{Becirevic:2018afm}.
\begin{figure}[!htb]
\begin{center}
 \includegraphics[width=0.7\hsize]{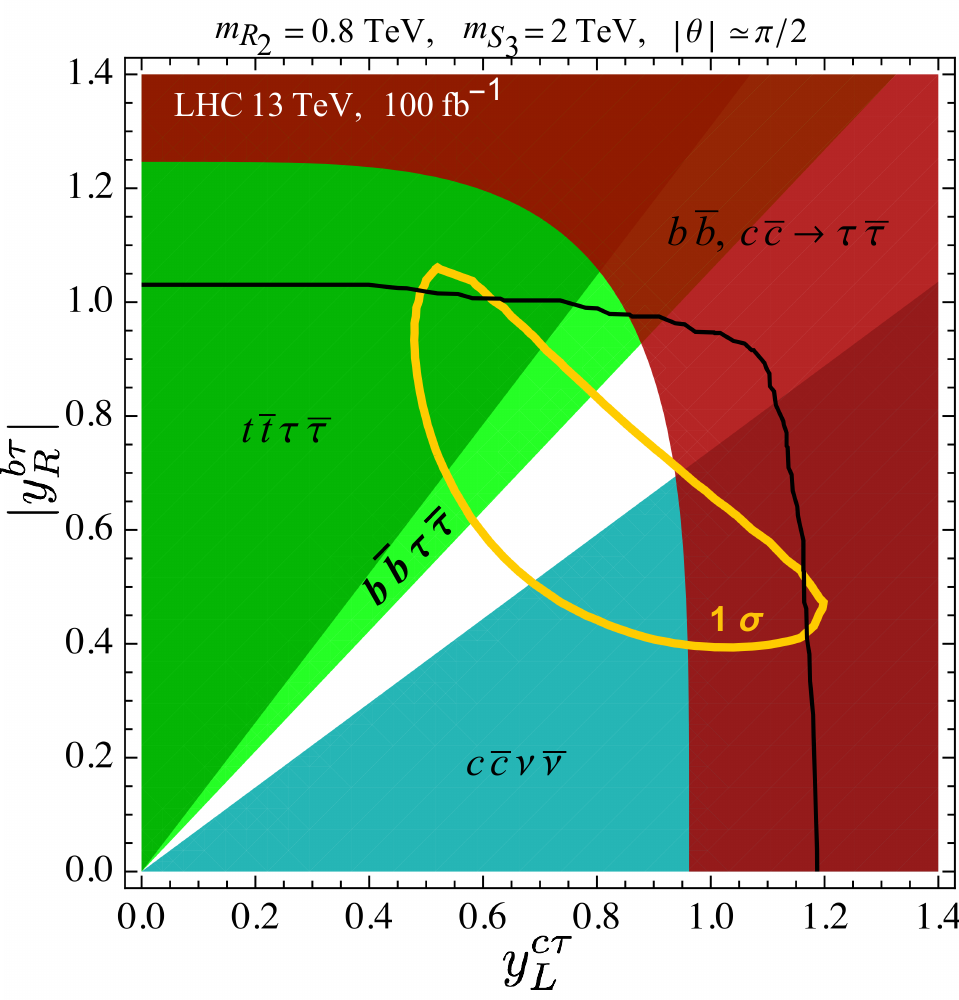}
 \vspace{-0.5cm}
\end{center}
\caption{Most important LHC limits for each LQ process at a projected luminosity of 100\,fb$^{-1}$. The red region is excluded by the high-$p_T$ di-tau search by ATLAS~\cite{Aaboud:2017sjh}, the green and turquoise exclusion regions come from LQ pair production searches by CMS~\cite{Sirunyan:2017yrk,CMS:2017kmd,Sirunyan:2018nkj}. The region of Yukawa couplings above the black line is excluded due to their non-perturbative values below the GUT scale~(see \cite{Becirevic:2018afm} for more details). The yellow contour denotes the $1\,\sigma$ region of the fit to the low-energy observables.}
\label{fig:LHCbound}
\end{figure}

\section{Summary and outlook}
Hints of lepton universality violation inconsistency in $\rk$ and $\rd$ with the SM have trigerred a gold rush in the flavour community that resulted in many proposed models. Leptoquarks are one possibility which are probably phenomenologically most suited to the observed puzzles which all dwell in leptoquarks' natural habitat of semileptonic processes. The $U_1$ vector leptoquark accommodates all observed lepton universality anomalies and has to be accompanyed with its own mass generation mechanism stemming from the UV.
We have proposed a two scalar LQ model that accommodates the observed LFU ratios in $B$-meson decays and is compatible with other low energy constraints as well as with direct searches at the LHC. The model has an $SU(5)$ origin that relates Yukawa couplings of the two LQs through a mixing angle. The model remains perturbative up to the unification scale. We propose signals of the two light LQs at the LHC and spell out predictions for several flavour observables. We predict and correlate $\mathcal{B}(B\to K\mu\tau)$ with $\mathcal{B}(B\to K ^{(\ast)} \nu \nu)$, as well as derive a lower bound for $\mathcal{B}(\tau\to\mu\gamma)$, which should be in reach of the Belle 2 experiment.


\begin{thebibliography}{99}


  \bibitem{Amhis:2019ckw} 
  Y.~S.~Amhis {\it et al.} [HFLAV Collaboration],
  arXiv:1909.12524 [hep-ex].


\bibitem{1205.5442} 
  J.~P.~Lees {\it et al.} [BaBar Collaboration],
  Phys.\ Rev.\ Lett.\  {\bf 109}, 101802 (2012)
  doi:10.1103/PhysRevLett.109.101802
  [arXiv:1205.5442 [hep-ex]].


\bibitem{1303.0571} 
  J.~P.~Lees {\it et al.} [BaBar Collaboration],
  Phys.\ Rev.\ D {\bf 88}, no. 7, 072012 (2013)
  doi:10.1103/PhysRevD.88.072012
  [arXiv:1303.0571 [hep-ex]].


\bibitem{1507.03233} 
  M.~Huschle {\it et al.} [Belle Collaboration],
  Phys.\ Rev.\ D {\bf 92}, no. 7, 072014 (2015)
  doi:10.1103/PhysRevD.92.072014
  [arXiv:1507.03233 [hep-ex]].


\bibitem{1506.08614} 
  R.~Aaij {\it et al.} [LHCb Collaboration],
  Phys.\ Rev.\ Lett.\  {\bf 115}, no. 11, 111803 (2015)
  Erratum: [Phys.\ Rev.\ Lett.\  {\bf 115}, no. 15, 159901 (2015)]
  doi:10.1103/PhysRevLett.115.159901, 10.1103/PhysRevLett.115.111803
  [arXiv:1506.08614 [hep-ex]].


\bibitem{1612.00529} 
  S.~Hirose {\it et al.} [Belle Collaboration],
  Phys.\ Rev.\ Lett.\  {\bf 118}, no. 21, 211801 (2017)
  doi:10.1103/PhysRevLett.118.211801
  [arXiv:1612.00529 [hep-ex]].


\bibitem{1709.00129} 
  S.~Hirose {\it et al.} [Belle Collaboration],
  Phys.\ Rev.\ D {\bf 97}, no. 1, 012004 (2018)
  doi:10.1103/PhysRevD.97.012004
  [arXiv:1709.00129 [hep-ex]].


\bibitem{1708.08856} 
  R.~Aaij {\it et al.} [LHCb Collaboration],
  Phys.\ Rev.\ Lett.\  {\bf 120}, no. 17, 171802 (2018)
  doi:10.1103/PhysRevLett.120.171802
  [arXiv:1708.08856 [hep-ex]].


\bibitem{1709.02505}
	*** Not Found with lookup: 'find eprint 1709.02505'

\bibitem{1904.08794} 
  A.~Abdesselam {\it et al.} [Belle Collaboration],
  arXiv:1904.08794 [hep-ex].


\bibitem{1606.08030} 
  D.~Bigi and P.~Gambino,
  Phys.\ Rev.\ D {\bf 94}, no. 9, 094008 (2016)
  doi:10.1103/PhysRevD.94.094008
  [arXiv:1606.08030 [hep-ph]].


\bibitem{1703.05330} 
  F.~U.~Bernlochner, Z.~Ligeti, M.~Papucci and D.~J.~Robinson,
  Phys.\ Rev.\ D {\bf 95}, no. 11, 115008 (2017)
  Erratum: [Phys.\ Rev.\ D {\bf 97}, no. 5, 059902 (2018)]
  doi:10.1103/PhysRevD.95.115008, 10.1103/PhysRevD.97.059902
  [arXiv:1703.05330 [hep-ph]].


\bibitem{1707.09509} 
  D.~Bigi, P.~Gambino and S.~Schacht,
  JHEP {\bf 1711}, 061 (2017)
  doi:10.1007/JHEP11(2017)061
  [arXiv:1707.09509 [hep-ph]].


\bibitem{1707.09977} 
  S.~Jaiswal, S.~Nandi and S.~K.~Patra,
  JHEP {\bf 1712}, 060 (2017)
  doi:10.1007/JHEP12(2017)060
  [arXiv:1707.09977 [hep-ph]].


\bibitem{Aaij:2019wad} 
  R.~Aaij {\it et al.} [LHCb Collaboration],
  Phys.\ Rev.\ Lett.\  {\bf 122}, no. 19, 191801 (2019)
  doi:10.1103/PhysRevLett.122.191801
  [arXiv:1903.09252 [hep-ex]].


\bibitem{1605.07633} 
  M.~Bordone, G.~Isidori and A.~Pattori,
  Eur.\ Phys.\ J.\ C {\bf 76}, no. 8, 440 (2016)
  doi:10.1140/epjc/s10052-016-4274-7
  [arXiv:1605.07633 [hep-ph]].


\bibitem{Aaij:2017vbb} 
  R.~Aaij {\it et al.} [LHCb Collaboration],
  JHEP {\bf 1708}, 055 (2017)
  doi:10.1007/JHEP08(2017)055
  [arXiv:1705.05802 [hep-ex]].


\bibitem{1505.03925} 
  H.~Na {\it et al.} [HPQCD Collaboration],
  Phys.\ Rev.\ D {\bf 92}, no. 5, 054510 (2015)
  Erratum: [Phys.\ Rev.\ D {\bf 93}, no. 11, 119906 (2016)]
  doi:10.1103/PhysRevD.93.119906, 10.1103/PhysRevD.92.054510
  [arXiv:1505.03925 [hep-lat]].


\bibitem{1503.07237} 
  J.~A.~Bailey {\it et al.} [MILC Collaboration],
  Phys.\ Rev.\ D {\bf 92}, no. 3, 034506 (2015)
  doi:10.1103/PhysRevD.92.034506
  [arXiv:1503.07237 [hep-lat]].


\bibitem{1311.5071} 
  M.~Atoui, D.~Becirevic, V.~Morénas and F.~Sanfilippo,
  PoS LATTICE {\bf 2013}, 384 (2014)
  doi:10.22323/1.187.0384
  [arXiv:1311.5071 [hep-lat]].


\bibitem{1612.07233} 
  Y.~Amhis {\it et al.} [HFLAV Collaboration],
  Eur.\ Phys.\ J.\ C {\bf 77}, no. 12, 895 (2017)
  doi:10.1140/epjc/s10052-017-5058-4
  [arXiv:1612.07233 [hep-ex]].


\bibitem{1806.10155} 
  F.~Feruglio, P.~Paradisi and O.~Sumensari,
  JHEP {\bf 1811}, 191 (2018)
  doi:10.1007/JHEP11(2018)191
  [arXiv:1806.10155 [hep-ph]].


\bibitem{1811.09603} 
  M.~Blanke, A.~Crivellin, S.~de Boer, T.~Kitahara, M.~Moscati, U.~Nierste and I.~Nišandžić,
  Phys.\ Rev.\ D {\bf 99}, no. 7, 075006 (2019)
  doi:10.1103/PhysRevD.99.075006
  [arXiv:1811.09603 [hep-ph]].


\bibitem{1808.08179} 
  A.~Angelescu, D.~Bečirević, D.~A.~Faroughy and O.~Sumensari,
  JHEP {\bf 1810}, 183 (2018)
  doi:10.1007/JHEP10(2018)183
  [arXiv:1808.08179 [hep-ph]].


\bibitem{1506.08896} 
  M.~Freytsis, Z.~Ligeti and J.~T.~Ruderman,
  Phys.\ Rev.\ D {\bf 92}, no. 5, 054018 (2015)
  doi:10.1103/PhysRevD.92.054018
  [arXiv:1506.08896 [hep-ph]].


\bibitem{1903.03102} 
  A.~Abdesselam {\it et al.} [Belle Collaboration],
  arXiv:1903.03102 [hep-ex].


\bibitem{1606.03164} 
  A.~K.~Alok, D.~Kumar, S.~Kumbhakar and S.~U.~Sankar,
  Phys.\ Rev.\ D {\bf 95}, no. 11, 115038 (2017)
  doi:10.1103/PhysRevD.95.115038
  [arXiv:1606.03164 [hep-ph]].


\bibitem{1811.08899} 
  S.~Iguro, T.~Kitahara, Y.~Omura, R.~Watanabe and K.~Yamamoto,
  JHEP {\bf 1902}, 194 (2019)
  doi:10.1007/JHEP02(2019)194
  [arXiv:1811.08899 [hep-ph]].


\bibitem{1611.06676} 
  R.~Alonso, B.~Grinstein and J.~Martin Camalich,
  Phys.\ Rev.\ Lett.\  {\bf 118}, no. 8, 081802 (2017)
  doi:10.1103/PhysRevLett.118.081802
  [arXiv:1611.06676 [hep-ph]].


\bibitem{1708.04072} 
  A.~G.~Akeroyd and C.~H.~Chen,
  Phys.\ Rev.\ D {\bf 96}, no. 7, 075011 (2017)
  doi:10.1103/PhysRevD.96.075011
  [arXiv:1708.04072 [hep-ph]].


\bibitem{1605.09308} 
  X.~Q.~Li, Y.~D.~Yang and X.~Zhang,
  JHEP {\bf 1608}, 054 (2016)
  doi:10.1007/JHEP08(2016)054
  [arXiv:1605.09308 [hep-ph]].


\bibitem{1606.00524} 
  F.~Feruglio, P.~Paradisi and A.~Pattori,
  Phys.\ Rev.\ Lett.\  {\bf 118}, no. 1, 011801 (2017)
  doi:10.1103/PhysRevLett.118.011801
  [arXiv:1606.00524 [hep-ph]].


\bibitem{1705.00929} 
  F.~Feruglio, P.~Paradisi and A.~Pattori,
  JHEP {\bf 1709}, 061 (2017)
  doi:10.1007/JHEP09(2017)061
  [arXiv:1705.00929 [hep-ph]].


\bibitem{1903.09578} 
  M.~Algueró, B.~Capdevila, A.~Crivellin, S.~Descotes-Genon, P.~Masjuan, J.~Matias and J.~Virto,
  Eur.\ Phys.\ J.\ C {\bf 79}, no. 8, 714 (2019)
  doi:10.1140/epjc/s10052-019-7216-3
  [arXiv:1903.09578 [hep-ph]].


\bibitem{1706.01868} 
  L.~Di Luzio and M.~Nardecchia,
  Eur.\ Phys.\ J.\ C {\bf 77}, no. 8, 536 (2017)
  doi:10.1140/epjc/s10052-017-5118-9
  [arXiv:1706.01868 [hep-ph]].


\bibitem{1706.07808} 
  D.~Buttazzo, A.~Greljo, G.~Isidori and D.~Marzocca,
  JHEP {\bf 1711}, 044 (2017)
  doi:10.1007/JHEP11(2017)044
  [arXiv:1706.07808 [hep-ph]].


\bibitem{1712.01368} 
  M.~Bordone, C.~Cornella, J.~Fuentes-Martin and G.~Isidori,
  Phys.\ Lett.\ B {\bf 779}, 317 (2018)
  doi:10.1016/j.physletb.2018.02.011
  [arXiv:1712.01368 [hep-ph]].


\bibitem{1805.09328} 
  M.~Bordone, C.~Cornella, J.~Fuentes-Martín and G.~Isidori,
  JHEP {\bf 1810}, 148 (2018)
  doi:10.1007/JHEP10(2018)148
  [arXiv:1805.09328 [hep-ph]].


\bibitem{1903.11517} 
  C.~Cornella, J.~Fuentes-Martin and G.~Isidori,
  JHEP {\bf 1907}, 168 (2019)
  doi:10.1007/JHEP07(2019)168
  [arXiv:1903.11517 [hep-ph]].


\bibitem{1802.04274} 
  A.~Greljo and B.~A.~Stefanek,
  Phys.\ Lett.\ B {\bf 782}, 131 (2018)
  doi:10.1016/j.physletb.2018.05.033
  [arXiv:1802.04274 [hep-ph]].


\bibitem{1704.05835} 
  D.~Bečirević and O.~Sumensari,
  JHEP {\bf 1708}, 104 (2017)
  doi:10.1007/JHEP08(2017)104
  [arXiv:1704.05835 [hep-ph]].


\bibitem{1805.04917} 
  J.~E.~Camargo-Molina, A.~Celis and D.~A.~Faroughy,
  Phys.\ Lett.\ B {\bf 784}, 284 (2018)
  doi:10.1016/j.physletb.2018.07.051
  [arXiv:1805.04917 [hep-ph]].


\bibitem{1703.09226} 
  A.~Crivellin, D.~Müller and T.~Ota,
  JHEP {\bf 1709}, 040 (2017)
  doi:10.1007/JHEP09(2017)040
  [arXiv:1703.09226 [hep-ph]].


\bibitem{Becirevic:2018afm} 
  D.~Bečirević, I.~Doršner, S.~Fajfer, N.~Košnik, D.~A.~Faroughy and O.~Sumensari,
  Phys.\ Rev.\ D {\bf 98}, no. 5, 055003 (2018)
  doi:10.1103/PhysRevD.98.055003
  [arXiv:1806.05689 [hep-ph]].


\bibitem{Dorsner:2017wwn} 
  I.~Doršner, S.~Fajfer and N.~Košnik,
  Eur.\ Phys.\ J.\ C {\bf 77}, no. 6, 417 (2017)
  doi:10.1140/epjc/s10052-017-4987-2
  [arXiv:1701.08322 [hep-ph]].


\bibitem{Dorsner:2017ufx} 
  I.~Doršner, S.~Fajfer, D.~A.~Faroughy and N.~Košnik,
  JHEP {\bf 1710}, 188 (2017)
  doi:10.1007/JHEP10(2017)188
  [arXiv:1706.07779 [hep-ph]].


\bibitem{Grygier:2017tzo} 
  J.~Grygier {\it et al.} [Belle Collaboration],
  Phys.\ Rev.\ D {\bf 96}, no. 9, 091101 (2017)
  Addendum: [Phys.\ Rev.\ D {\bf 97}, no. 9, 099902 (2018)]
  doi:10.1103/PhysRevD.97.099902, 10.1103/PhysRevD.96.091101
  [arXiv:1702.03224 [hep-ex]].


\bibitem{ALEPH:2005ab} 
  S.~Schael {\it et al.} [ALEPH and DELPHI and L3 and OPAL and SLD Collaborations and LEP Electroweak Working Group and SLD Electroweak Group and SLD Heavy Flavour Group],
  Phys.\ Rept.\  {\bf 427}, 257 (2006)
  doi:10.1016/j.physrep.2005.12.006
  [hep-ex/0509008].


\bibitem{Dekens:2018bci} 
  W.~Dekens, J.~de Vries, M.~Jung and K.~K.~Vos,
  JHEP {\bf 1901}, 069 (2019)
  doi:10.1007/JHEP01(2019)069
  [arXiv:1809.09114 [hep-ph]].


\bibitem{1811.07920} 
  A.~Greljo, J.~Martin Camalich and J.~D.~Ruiz-Álvarez,
  Phys.\ Rev.\ Lett.\  {\bf 122}, no. 13, 131803 (2019)
  doi:10.1103/PhysRevLett.122.131803
  [arXiv:1811.07920 [hep-ph]].


\bibitem{Aaboud:2017sjh} 
  M.~Aaboud {\it et al.} [ATLAS Collaboration],
  JHEP {\bf 1801}, 055 (2018)
  doi:10.1007/JHEP01(2018)055
  [arXiv:1709.07242 [hep-ex]].


\bibitem{Sirunyan:2017yrk} 
  A.~M.~Sirunyan {\it et al.} [CMS Collaboration],
  JHEP {\bf 1707}, 121 (2017)
  doi:10.1007/JHEP07(2017)121
  [arXiv:1703.03995 [hep-ex]].


\bibitem{CMS:2017kmd} 
  CMS Collaboration [CMS Collaboration],
  CMS-PAS-SUS-16-036.


\bibitem{Sirunyan:2018nkj} 
  A.~M.~Sirunyan {\it et al.} [CMS Collaboration],
  Eur.\ Phys.\ J.\ C {\bf 78}, 707 (2018)
  doi:10.1140/epjc/s10052-018-6143-z
  [arXiv:1803.02864 [hep-ex]].




  

\bibitem{Amhis:2019ckw} 
  Y.~S.~Amhis {\it et al.} [HFLAV Collaboration],
  arXiv:1909.12524 [hep-ex].


\bibitem{1205.5442} 
  J.~P.~Lees {\it et al.} [BaBar Collaboration],
  Phys.\ Rev.\ Lett.\  {\bf 109}, 101802 (2012)
  doi:10.1103/PhysRevLett.109.101802
  [arXiv:1205.5442 [hep-ex]].


\bibitem{1303.0571} 
  J.~P.~Lees {\it et al.} [BaBar Collaboration],
  Phys.\ Rev.\ D {\bf 88}, no. 7, 072012 (2013)
  doi:10.1103/PhysRevD.88.072012
  [arXiv:1303.0571 [hep-ex]].


\bibitem{1507.03233} 
  M.~Huschle {\it et al.} [Belle Collaboration],
  Phys.\ Rev.\ D {\bf 92}, no. 7, 072014 (2015)
  doi:10.1103/PhysRevD.92.072014
  [arXiv:1507.03233 [hep-ex]].


\bibitem{1506.08614} 
  R.~Aaij {\it et al.} [LHCb Collaboration],
  Phys.\ Rev.\ Lett.\  {\bf 115}, no. 11, 111803 (2015)
  Erratum: [Phys.\ Rev.\ Lett.\  {\bf 115}, no. 15, 159901 (2015)]
  doi:10.1103/PhysRevLett.115.159901, 10.1103/PhysRevLett.115.111803
  [arXiv:1506.08614 [hep-ex]].


\bibitem{1612.00529} 
  S.~Hirose {\it et al.} [Belle Collaboration],
  Phys.\ Rev.\ Lett.\  {\bf 118}, no. 21, 211801 (2017)
  doi:10.1103/PhysRevLett.118.211801
  [arXiv:1612.00529 [hep-ex]].


\bibitem{1709.00129} 
  S.~Hirose {\it et al.} [Belle Collaboration],
  Phys.\ Rev.\ D {\bf 97}, no. 1, 012004 (2018)
  doi:10.1103/PhysRevD.97.012004
  [arXiv:1709.00129 [hep-ex]].


\bibitem{1708.08856} 
  R.~Aaij {\it et al.} [LHCb Collaboration],
  Phys.\ Rev.\ Lett.\  {\bf 120}, no. 17, 171802 (2018)
  doi:10.1103/PhysRevLett.120.171802
  [arXiv:1708.08856 [hep-ex]].


\bibitem{1709.02505}
	*** Not Found with lookup: 'find eprint 1709.02505'

\bibitem{1904.08794} 
  A.~Abdesselam {\it et al.} [Belle Collaboration],
  arXiv:1904.08794 [hep-ex].


\bibitem{1606.08030} 
  D.~Bigi and P.~Gambino,
  Phys.\ Rev.\ D {\bf 94}, no. 9, 094008 (2016)
  doi:10.1103/PhysRevD.94.094008
  [arXiv:1606.08030 [hep-ph]].


\bibitem{1703.05330} 
  F.~U.~Bernlochner, Z.~Ligeti, M.~Papucci and D.~J.~Robinson,
  Phys.\ Rev.\ D {\bf 95}, no. 11, 115008 (2017)
  Erratum: [Phys.\ Rev.\ D {\bf 97}, no. 5, 059902 (2018)]
  doi:10.1103/PhysRevD.95.115008, 10.1103/PhysRevD.97.059902
  [arXiv:1703.05330 [hep-ph]].


\bibitem{1707.09509} 
  D.~Bigi, P.~Gambino and S.~Schacht,
  JHEP {\bf 1711}, 061 (2017)
  doi:10.1007/JHEP11(2017)061
  [arXiv:1707.09509 [hep-ph]].


\bibitem{1707.09977} 
  S.~Jaiswal, S.~Nandi and S.~K.~Patra,
  JHEP {\bf 1712}, 060 (2017)
  doi:10.1007/JHEP12(2017)060
  [arXiv:1707.09977 [hep-ph]].


\bibitem{Aaij:2019wad} 
  R.~Aaij {\it et al.} [LHCb Collaboration],
  Phys.\ Rev.\ Lett.\  {\bf 122}, no. 19, 191801 (2019)
  doi:10.1103/PhysRevLett.122.191801
  [arXiv:1903.09252 [hep-ex]].


\bibitem{1605.07633} 
  M.~Bordone, G.~Isidori and A.~Pattori,
  Eur.\ Phys.\ J.\ C {\bf 76}, no. 8, 440 (2016)
  doi:10.1140/epjc/s10052-016-4274-7
  [arXiv:1605.07633 [hep-ph]].


\bibitem{Aaij:2017vbb} 
  R.~Aaij {\it et al.} [LHCb Collaboration],
  JHEP {\bf 1708}, 055 (2017)
  doi:10.1007/JHEP08(2017)055
  [arXiv:1705.05802 [hep-ex]].


\bibitem{1505.03925} 
  H.~Na {\it et al.} [HPQCD Collaboration],
  Phys.\ Rev.\ D {\bf 92}, no. 5, 054510 (2015)
  Erratum: [Phys.\ Rev.\ D {\bf 93}, no. 11, 119906 (2016)]
  doi:10.1103/PhysRevD.93.119906, 10.1103/PhysRevD.92.054510
  [arXiv:1505.03925 [hep-lat]].


\bibitem{1503.07237} 
  J.~A.~Bailey {\it et al.} [MILC Collaboration],
  Phys.\ Rev.\ D {\bf 92}, no. 3, 034506 (2015)
  doi:10.1103/PhysRevD.92.034506
  [arXiv:1503.07237 [hep-lat]].


\bibitem{1311.5071} 
  M.~Atoui, D.~Becirevic, V.~Morénas and F.~Sanfilippo,
  PoS LATTICE {\bf 2013}, 384 (2014)
  doi:10.22323/1.187.0384
  [arXiv:1311.5071 [hep-lat]].


\end{thebibliography}
\end{document}